\documentclass[10pt]{iopart}

\usepackage{iopams}  
\usepackage{graphicx}
\begin{document}

\title[~]{The perturbative QCD
factorization approach in high energy nuclear collisions}

\author{Ivan Vitev\dag\ 
\footnote[3]{E-mail address: {\tt ivitev@lanl.gov} }
}

\address{\dag\ Los Alamos National Laboratory, Mail Stop H846,
Los Alamos, NM 87545, USA}

\begin{abstract}

I discuss the systematic modifications to the perturbative QCD 
factorization approach in high energy  $\ell+A$,  $p+A$ and $A+A$ 
reactions. These include transverse momentum diffusion manifest 
in the Cronin effect and a small increase in the dijet acoplanarity; 
nuclear size enhanced power corrections that lead to shadowing 
in deeply inelastic scattering and suppression of single and 
double inclusive hadron production at forward rapidity at RHIC 
but disappear as a function of the transverse momentum; inelastic 
attenuation of the jet cross sections or jet quenching that 
persists to much higher $p_T$.

\end{abstract}


\submitto{\JPG}

\section{Introduction}

The interpretation~\cite{Exp,Thy} of the copious experimental results
from the past, present and future heavy ion programs necessitates 
a reliable theoretical framework for the calculation of a 
variety of moderate and large transverse momentum processes,
$Q^2 \sim p_T^2 \gg \Lambda_{\rm QCD}^2$,
that can be systematically  extended to incorporate corrections arising 
from the many body nuclear dynamics. Such theoretical framework is the 
perturbative QCD factorization approach~\cite{Collins:gx}, 
a direct generalization of the naive parton model~\cite{Bjorken:1969ja}. 
According to the factorization theorem, the observable hadronic cross 
sections can be expressed as  
\begin{eqnarray}
 \sigma_{\rm hadronic} & = &  \sigma_{\rm partonic} 
\left( x_i, z_j;  \mu_r; \mu_{f\, i}, \mu_{d\, j} \right) 
\; \otimes  \; \left\{ \prod_i  \phi_{i/h_i}(x_i, \mu_{f\, i})  \right\} 
 \nonumber \\
&& \; \otimes  \;  \left\{ \prod_j D_{h_j/j}(z_j, \mu_{d\, j}) \right\} ;, 
\label{factorization}
\end{eqnarray}
where  $\otimes$ denotes the standard convolution over 
the internal kinematic variables of the reaction. 
In Eq.~(\ref{factorization}) $\mu_{r},\;\mu_{f\, i}$ and   
$\mu_{d\, j}$ are the renormalization, factorization and 
fragmentation scales, respectively.  $\phi_{i/h_i}(x_i, \mu_{f\, i})$ 
is the distribution function (PDF) of parton ``$i$'' in the hadron  $h_i$ 
and  $D_{h_j/j}(z_j, \mu_{d\, j})$ is the fragmentation or decay 
function (FF) of parton ``$j$'' into hadron $h_j$. Factorization 
not only separates the short- and long-distance QCD dynamics but 
implies universality of the PDFs and FFs and infrared safety of the 
hard scattering partonic cross sections. If the $k_{T \, i}$ and/or the 
$j_{T\, i}$ dependence of the PDFs,  FFs and $\sigma_{\rm partonic}$  
is kept explicit, one arrives at a $k_T$ and/or $j_T$ factorized 
form~\cite{Collins:1981uw}. At present, limited statistics hampers 
any reliable extraction of such transverse degrees of freedom. 
Integrating $k_{T \, i}$ and $j_{T\, i}$  out leads to the most 
commonly employed double collinear limit of QCD.

Factorization has been discussed in detail for  
a limited number of processes. Explicit examples of some fundamental 
cross sections  are given  below  to lowest order and leading twist. 
All scales are suppressed and the definitions of the kinematic variables 
can be found in~\cite{Brock:1993sz}. 
These examples include:
\begin{enumerate}     
\item Electron-positron annihilation into hadrons~\cite{Collins:1981uk}:
\begin{equation}
\frac{d \sigma_h}{dx dy}  = 
N_c \frac{ \pi \alpha_{em}^2 }{q^2} \sum_j Q_i^2 
\left( 1 + \cos^2 \theta_{\rm cm} \right) D_{h_j/j}(z_j) \;.
\label{epemannih}
\end{equation}
Here $x = 2 p \cdot q / q^2$,  
$y = \frac{1}{2} (1 - \cos \theta_{\rm cm} )$ and $Q_i$ 
is the fractional electric charge of (anti)quarks in
units of $e$.    
For a discussion on the structure of  the near-side correlation 
function from fragmentation  in $e^+ + e^-$  
annihilation see~\cite{Majumder:2004wh}.

\item  Deeply inelastic lepton-hadron scattering (DIS) with the 
longitudinal and transverse structure  functions:
\begin{equation} 
F_T = \frac{1}{2} \sum_i Q_i^2\,  \phi_{f}(x) \; , 
\label{strfun}
\qquad  F_L  =  0 \;.  
\end{equation} 
In Eq.~(\ref{strfun}) $x = x_B = -q^2 / 2 p \cdot q =  
Q^2 / 2 m_N \nu$ with $\nu = E - E^\prime$.
Recently the nuclear enhanced high twist corrections  
to the DIS cross sections have been
evaluated within the factorization approach~\cite{Qiu:2003vd,Qiu:2004qk}.
Similar results from final state multiple scattering have been
found in~\cite{Brodsky:1989qz}. 

\item The Drell-Yan process~\cite{Bodwin:1984hc}:  
\begin{equation}
\hspace*{-2cm} \frac{d \sigma_{l^+l^-}}{dq^2}  = \sum_i 
\phi_{i/N}(x_a) \phi_{\, \bar{i}/N^\prime}(x_b) \;
\frac{ 4 \pi \alpha_{em}^2  Q_i^2 }{3 N_c q^2}  \; 
\delta \left( q^2 -(x_a p + x_b p^\prime )^2 \right) \;.
\label{drellyan}
\end{equation}
In Eqs.~(\ref{epemannih}) and (\ref{drellyan}) $N_c = 3$ 
is the number of colors. The momentum fractions 
$x_a = p_a^+/P_a^+$ and   $x_b = p_b^+/P_b^+$.

\item Inclusive hadron production in $N+N$  
collisions to leading power and leading power 
corrections~\cite{Ellis:1978sf}. The single and double
inclusive distributions accessible to ${\cal O}(\alpha_s^2)$ 
at leading twist read~\cite{Qiu:2004da}:  
\begin{eqnarray} 
\label{single}
\hspace*{-2cm}
\frac{ d\sigma^{h_1 }_{NN} }{ dy_1  d^2p_{T_1} }  
& = &  \sum_{abcd}
\int \frac{dz_1}{z_1^2} D_{h_1/c}(z_1) 
\int d x_a   \frac{\phi_{a/N}(x_a)}{x_a}  
\left[\frac{1}{x_a S +  {U}/{z_1} }\right] \nonumber \\  
&& \times \; \frac{ \alpha_s^2}{S} \;
\frac{\phi_{b/N}(x_b)}{x_b}\,  
|\overline {M}_{ab\rightarrow cd}|^2  \;, 
\\
\hspace*{-2cm}
\frac{ d\sigma^{h_1 h_2}_{NN} }
{ dy_1  dy_2 d^2p_{T_1}  d^2p_{T_2} } 
&=& 
\frac{\delta (\Delta \varphi - \pi)}{p_{T_1} p_{T_2} } 
\sum_{abcd} 
\int \frac{dz_1}{z_1} \, D_{h_1/c}(z_1) \,
D_{h_2/d} (z_2)\,  \frac{\phi_{a/N}(\bar{x}_a)}{\bar{x}_a} \,
\nonumber \\ 
&& \times \;  \frac{\alpha_s^2}{{S}^2 }
\; \frac{\phi_{b/N}(x_b)}{x_b}\,  
|\overline {M}_{ab\rightarrow cd}|^2  \;.
\label{double}
\end{eqnarray}
In Eqs.~(\ref{single}) and (\ref{double}) $S,\; T$ and $U$
are the {\em hadronic} Mandelstam variables~\cite{Qiu:2004da}.  
The dependence on the small momentum fraction $x_b$, relevant 
to forward rapidity studies at RHIC,  is explicitly isolated.

\item  Heavy quark and heavy quark bound state 
production~\cite{Bodwin:1994jh}.

\end{enumerate}

\begin{figure}[t!]
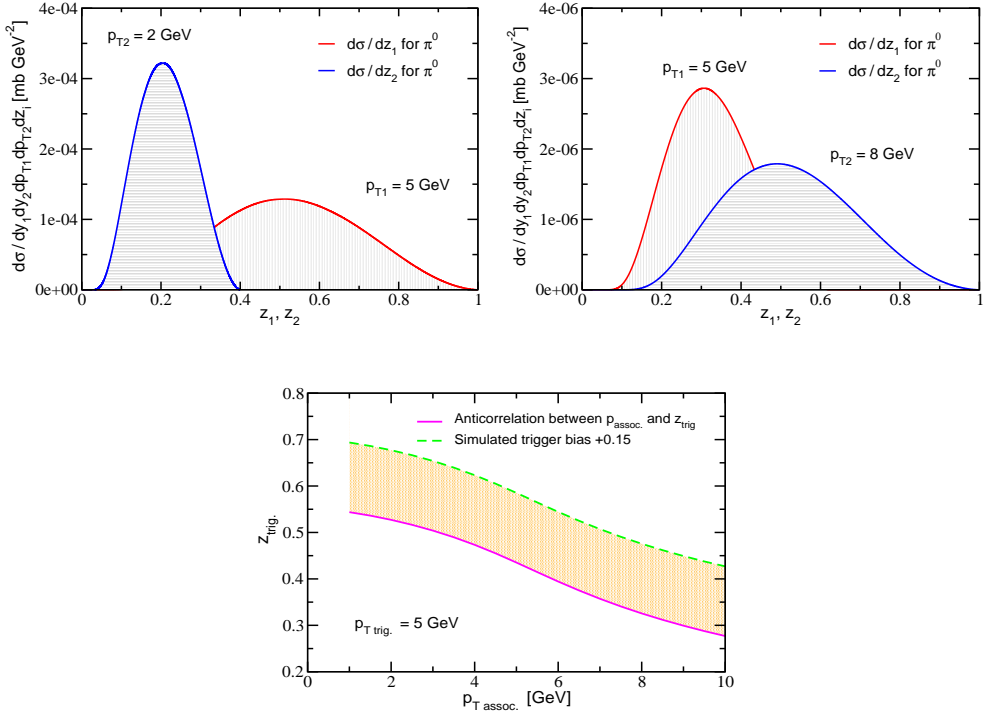

\vspace*{.4cm}
\includegraphics[width=2.5in,height=1.7in]{Fseesaw-0.eps}
\hspace*{0.2cm}\includegraphics[width=2.5in,height=1.7in]{Fseesaw-1.eps}
\begin{center}
\vspace*{.6cm}
\includegraphics[width=2.5in,height=1.7in]{Fseesaw-3.eps}
\end{center} 
\caption{ Left panel: 
the differential double inclusive $\pi^0$ distributions versus
the fragmentation momentum fractions  $z_1,\, z_2$  for a trigger
pion of $p_T = 5$~GeV and associated pion of  $p_T = 2$~GeV.
Right panel: the same cross sections  for a trigger
pion of $p_T = 5$~GeV and associated pion of $p_T = 8$~GeV. 
Bottom panel: the anti-correlation between $z_{\rm trig}$ and 
$p_{T\; assoc}$ for a fixed  $p_{T\; {\rm trig}} = 5$~GeV. The dashed 
line simulates a trigger bias $\Delta z_{\rm trig} = 0.15$. }  
\label{fig1-2}
\end{figure}

\subsection{The fragmentation seesaw  analogy}

One of the exciting theoretical questions, prompted 
by the recent experimental results~\cite{Adcox:2001mf,expPtoPI},
is about the nature of the hadron production 
mechanism at small and moderate $p_T$ in $A+A$ 
reactions~\cite{Gribov:1984tu,Vitev:2001zn,Fries:2004ej,Hirano:2004er}. 
The measurement of correlations~\cite{Adler:2002tq} in $p+p$ and 
especially $d+Au$ collisions already provide sufficient
evidence that back-to-back hard partonic scattering and fragmentation,
Eqs.~(\ref{single}) and (\ref{double}), 
give a dominant contribution to the particle cross sections. 
Corrections in the basic perturbative formulas arise both from 
higher orders in $\alpha_s$ and  from higher twist. 
These may be enhanced by the nuclear size and the 
density of the quark-gluon plasma (QGP) in the deconfined 
phase. Such corrections can be systematically 
organized in the framework of the factorization approach 
as follows~\cite{Qiu:2003cg}:
\begin{eqnarray}
\sigma_{\rm hadron} & = & \sigma_0^{(2)} 
\left( 1 + C^{(2)}_1 \alpha_s + C^{(2)}_2  \alpha_s^2 
+   C^{(2)}_3  \alpha_s^3 + \cdots  \right) T^{(2)}
 \nonumber  \\
&& + \frac{ \sigma_0^{(4)} }{Q^2}  
\left( 1 + C^{(2)}_1 \alpha_s + C^{(2)}_2  \alpha_s^2 
+   C^{(2)}_3  \alpha_s^3 + \cdots  \right) T^{(4)} 
  \nonumber  \\
&& + \frac{ \sigma_0^{(6)} }{Q^4} 
 \left( 1 + C^{(2)}_1 \alpha_s + C^{(2)}_2  \alpha_s^2 
+   C^{(2)}_3  \alpha_s^3 + \cdots  \right)  T^{(6)}  
 +  \cdots \;\; \quad
\label{series}
\end{eqnarray}     
In Eq.~(\ref{series}) $T^{(i)}$ are the twist ``i'' 
non-perturbative matrix elements. Odd-twist matrix
elements play an important role in spin 
physics~\cite{Qiu:1991pp} but are here neglected for simplicity.

As a test of the dominance of the lowest order, 
leading twist term one can experimentally look for a seesaw 
fragmentation {\em analogy}. The seesaw 
mechanism~\cite{Slansky:1981yr} extends the 
Standard Model~\cite{Weinberg:1967tq} by including a massive 
right-handed Majorana neutrino. 
The physical neutrino masses are given by 
$M_\nu = M_T M_R^{-1} M_T^\dagger$. When 
$M_R$ is very large and hence $\nu_R$ is unobservable, 
$M_\nu$ can become very small.

Similar  anti-correlation is present in Eq.~(\ref{double}). 
Naively, one might expect that a high-$p_T$ hadron trigger 
in the near side will fix the fragmentation momentum 
fraction $z_{\rm trig}$. 
In contrast,  $\delta \left( {p_{T1}}/{z_1} - {p_{T2}}/{z_2}  \right)$
implies  that  $z_{\rm trig}$ is inversely proportional to 
$p_{T \; \rm assoc}$. Figure~\ref{fig1-2} shows the  differential 
distributions  $ d\sigma^{h_1 h_2} / dy_1dy_2 dp_{T1} dp_{T2} dz_i$
for different choices of $p_{T \;  \rm trig}$ and  $p_{T \; \rm assoc}$. 
Experimental verification of the anti-correlation  presented 
in the bottom panel of Fig.~\ref{fig1-2} can  provide critical 
additional evidence in support of the dominance of the 
perturbative QCD  hadron production mechanism and the cross
section hierarchy summarized by Eq.~(\ref{series}).

\section{Transverse momentum diffusion}

Hard processes in the nuclear environment 
involve multiple parton interactions before or after the  
large $Q^2$ collision that are sensitive to the 
properties of the nuclear matter. In this section we 
consider the multiple elastic and 
incoherent scattering of energetic quarks and gluons
that penetrate hot and cold QCD matter. Traverse momentum
diffusion reflects the  $ \langle p_T^2 \rangle$-kick per unit 
length in the medium, given by the transport coefficient 
$\mu^2/\lambda_{q,g}$, and is the strong interaction dynamics
equivalent of the Moliere multiple scattering in 
QED.

Approximate solutions for the 
parton broadening can be obtained to leading power
and leading power corrections in the large ``$+$'' lightcone
momentum $p^+$~\cite{Qiu:2003pm}.  
For the instructive case of a normalized forward monochromatic beam 
($p_0=p_\parallel \equiv P$) the initial quark or gluon distribution
reads:
\begin{equation}  
\frac{d^3N^i}{d p^+ d^2 {\bf p}_\perp } |_{p^- =  
\frac{{\bf p}_\perp^2}{2 p^+ } }= 
\delta(p^+ - \sqrt{2} P) \delta^2({\bf p}_\perp )  \;.
\label{fbeam}
\end{equation}  
In Eq.~(\ref{fbeam}) the constraint on the parton's ``$-$'' lightcone 
component  from the $p^2=0$ on-shell condition is also shown. 
In the small angle approximation, rescattering leads to an
approximately Gaussian form~\cite{Qiu:2003pm}:  
\begin{equation}
\hspace*{-1cm}\frac{d^3N^{f}(p^+,{\bf p}_\perp)}{d p^+ d^2 {\bf p}_\perp }
|_{p^- =  \frac{ {\bf p}_\perp^2}{2 p^+ } }
= \frac{1}{2\pi} \frac{e^{-\frac{{\bf p}^2_\perp}{2\,\chi \,\mu^2 \xi}}}
 {\chi\, \mu^2\, \xi } \,  \delta\left[p^+ - 
\left(\sqrt{2}P-\frac{1}{\sqrt{2}} \frac{\chi \mu^2 \xi}{2 P}  
 \right) \right]  \;\;,
\label{gauss}  
\end{equation}
where $\chi = L/\lambda$  is the opacity or mean number of scatterings
and $\xi = {\cal O}(1)$. It should be realized that this form is 
not expected to describe well the large angle scattering. 
The tails of  the large  ${\bf p}_\perp$ distributions are 
expected to be  power-law like and die out at a much slower rate than the 
Gaussian  in Eq.~(\ref{gauss}).  This may lead to 
uncertainties in the separation of the jet and the underlying 
event~\cite{Adler:2002tq}.

\subsection{Applications of transverse momentum diffusion}

From Eq.~(\ref{gauss}) it is easy to demonstrate that the  
the accumulated acoplanarity momentum squared by an 
energetic parton probes a line integral through the 
color charge density:
\begin{eqnarray}
\langle \Delta {\bf k}_T^2 \rangle   \approx 
  2 \xi \int dz \, \frac{\mu^2}{\lambda_{q,g}} 
 & = & 2 \xi \int dz \,  \frac{3  C_R \pi \alpha_s^2}{2} \rho^g(z) 
\nonumber \\[1ex] 
& = & \left\{ \begin{array}{ll}   2 \xi \,  
\frac{3  C_R \pi \alpha_s^2}{2} \, \rho^g \langle L \rangle \, , 
&  {\rm static} \\[1ex]
 2\xi \, \frac{3  C_R \pi \alpha_s^2}{2} 
\frac{1}{A_\perp} \frac{dN^{g}}{dy}
\,   \ln \frac{  \langle L  \rangle }{\tau_0}\, , & 1+1D  
\end{array}  \right. 
\label{broad}
\end{eqnarray}
In Eq.~(\ref{broad}) the factor 2 comes from 2D diffusion,  
and $\rho^g$ is the  effective gluon density. 
For the 1+1D Bjorken expansion scenario $A_\perp$ is the 
transverse area of the interaction region, $\tau_0$ is the initial 
equilibration time and  $dN^g/dy$ is the effective gluon rapidity
density. 

Dynamical nuclear-induced modification in multi-particle 
production can be studied through the ratio~\cite{Qiu:2004da}:
\begin{equation}
  R^{(n)}_{AB}  =  \frac{d\sigma^{h_1 \cdots h_n}_{AB} / 
dy_1 \cdots dy_n d^2p_{T_1} \cdots d^2p_{T_n}} 
{\langle N^{\rm coll}_{AB} \rangle\, d\sigma^{h_1 \cdots h_n}_{NN} / 
dy_1 \cdots dy_n d^2p_{T_1} \cdots d^2p_{T_n}} \; .
\label{multi}
\end{equation}
Centrality dependence is implicit in Eq.~(\ref{multi}). 
The first and possibly most sensitive indicator of multi-parton
dynamics in cold nuclear matter is the Cronin effect~\cite{Cronin:1974zm}
in the single inclusive hadron production in $p+A$ reactions, 
$R^{(1)}_{pA}({\bf p}_T)$.

\begin{figure}[t!]
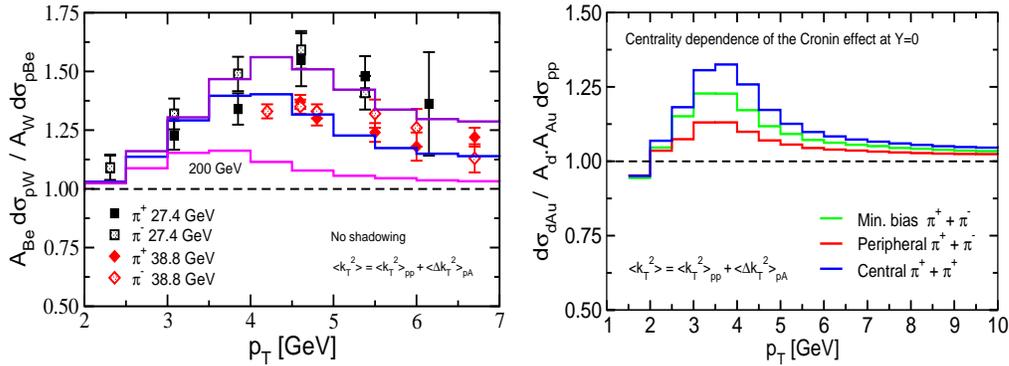

\vspace*{.2cm}
\includegraphics[width=2.6in,height=1.9in]{cron.SHnoSH.hq.eps}
\hspace*{0.2cm}
\vspace*{.2cm}
\includegraphics[width=2.5in,height=1.9in]{Cron.PI.cantral.eps} 
\vspace*{-.2cm}
\caption{ Left panel from~\cite{Vitev:2002pf}:
calculation of the Cronin effect for $\pi^+ + \pi^-$ 
at $\sqrt{s_{NN}} = 27.4$ and $38.8$~GeV  at midrapidity. 
The nuclear modification is identified from the ratio 
of the cross sections in  $p+W$ and $p+Be$ reactions.  
Right panel from~\cite{Vitev:2002pf}: predicted magnitude 
and  centrality dependence of $R_{dA}^{(1)}({\bf p}_T)$ at  
$\sqrt{s_{NN}} = 200$~GeV  and midrapidity at RHIC 
in the absence of significant  antishadowing.
}
\label{cronfig}
\end{figure}

Elastic transverse momentum diffusion at midrapidity 
is manifest in the Cronin  enhancement at $p_T \geq 1-2$~GeV 
and suppression for
 $p_T \leq 1-2$~GeV.  For a survey of theoretical models 
see~\cite{Accardi:2002ik}. Calculations of the
nuclear modification in low energy $p+A$ reactions employ 
$\mu^2/\lambda = 0.1 - 0.14$~GeV$^2$  and are shown in  the 
left hand side of Fig.~\ref{cronfig}. 
The predicted~\cite{Vitev:2002pf} centrality dependence of 
the Cronin effect in $d+Au$ reactions at $y=0$ 
and RHIC $\sqrt{s_{NN}}=200$~GeV is given in the 
right hand side of Fig.~\ref{cronfig} and confirmed by
experimental 
data~\cite{Adler:2002tq,d'Enterria:2004nv,Arsene:2003yk}.    
It is important to emphasize that even a small 
broadening may lead to a noticeable Cronin effect since 
it is amplified by the steepness of the partonic/hadronic
spectra.

It is interesting to look for other manifestations of transverse
momentum diffusion such as increased dijet acoplanarity, which 
will be reflected in the increased width of the away-side 
correlation function 
\begin{eqnarray}
C_2(\Delta \varphi) &=& \frac{1}{N_{\rm trig}} 
\frac{dN^{h_1h_2}_{\rm dijet}}{d\Delta \varphi}
 \nonumber \\  
& \approx & \frac{A_{\rm Near}}{\sqrt{2\pi} \sigma_{\rm Near} }  
e^{-\frac{\Delta \varphi^2}{2\sigma^2_{\rm Near}} } + 
\frac{A_{\rm Far}}{ \sqrt{2\pi} \sigma_{\rm Far} } 
e^{-\frac{(\Delta \varphi -\pi)^2}{2\sigma^2_{\rm Far}} }. \qquad 
\label{cor-fun}
\end{eqnarray}
In Eq.~(\ref{cor-fun}) the near-side width $\sigma_{\rm Near}$ of  
$C_2(\Delta \varphi)$ is determined by jet fragmentation and  
$\sigma_{\rm Far}$ reflects 
$\langle {\bf k}_T^2 \rangle_{\rm tot}  = \langle {\bf k}_T^2 \rangle_{pp} + 
\langle {\bf k}_T^2 \rangle_{nucl.}$.

\begin{figure}[t!]
\includegraphics[width=2.5in,height=2.3in]{Acoplanarity-kty.eps}
\hspace*{0.2cm}\includegraphics[width=2.6in,height=2.4in]{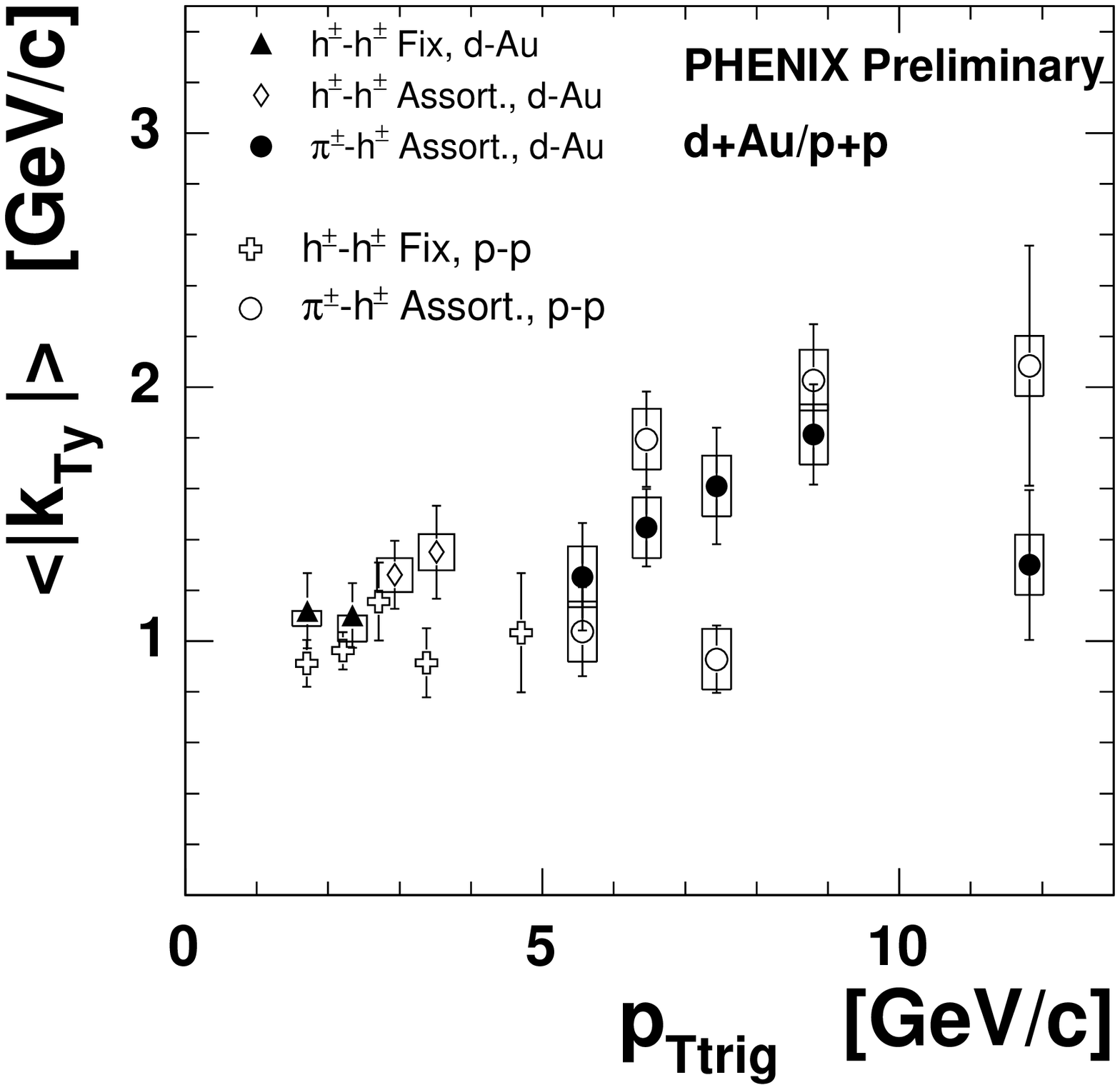}
\caption{ Left panel from~\cite{Qiu:2003pm}: the acoplanarity
momentum projection per parton $\langle | k_{T\,y} | \rangle $ 
and the corresponding width $\sigma_{\rm Far}$ of the away-side
correlation function for $p+p$ and $d+Au$. The approximations
of $z_{\rm trig} \rightarrow 1$ and constant 
$\langle | k_{T\,y} | \rangle_{pp}=0.75$ had been used.
Right panel from~\cite{Adler:2002tq}: preliminary PHENIX 
data on the vacuum and medium induced acoplanarity momentum 
projection per parton in cold nuclear matter for realistic 
finite $z_{ \rm trig}$. Note the absence of significant difference
between the $p+p$ and $d+Au$ measurements. }
\label{fig3-4}
\end{figure}

The left panel of Fig.~\ref{fig3-4} shows the upper limit of the 
medium induced acoplanarity 
relative to the vacuum $p+p$ result. The predicted increase
in the width of the away-side correlation function is small even
in the $z_{\rm trig} \rightarrow 1$ limit. For realistic 
fragmentation momentum fractions, see Fig.~\ref{fig1-2}, the 
experimentally observed enhancement of 
$\langle \Delta {\bf k}_T^2 \rangle_{\rm tot}$ should be significantly 
smaller. In addition, for dihadron angular correlations there 
are no amplification effects from the steepness of the 
underlying  parton spectra. The right panel of  Fig.~\ref{fig3-4}
indeed shows the lack of significant difference between 
$p+p$  and $d+Au$  measurements and certainly excludes 
monojet-based models.

\section{Nuclear enhanced power corrections}

\begin{figure}[t!]
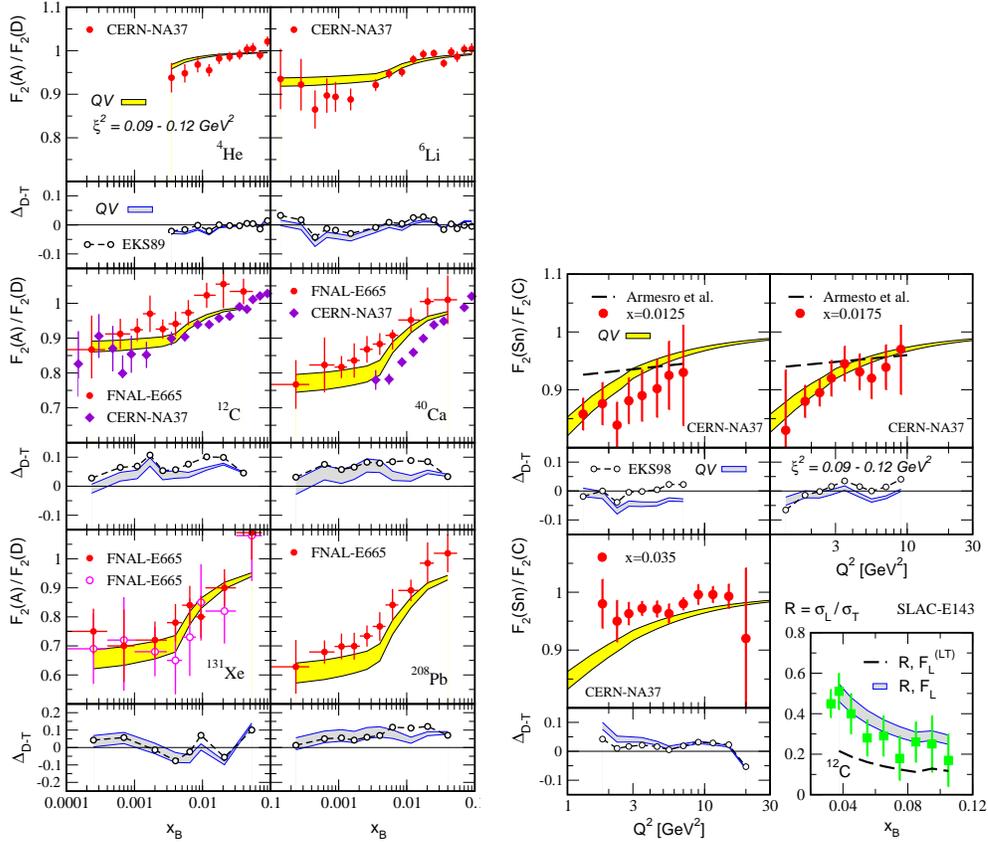

\includegraphics[width=2.5in,height=4.4in]{Fig-EMC-shadow.eps}
\hspace*{0.2cm}\includegraphics[width=2.5in,height=3.in]{Fig-FL-enhance.eps} 
\caption{ Left panel from~\cite{Qiu:2003vd}:
all-twist resummed $F_2(A)/F_2(D)$ calculation from 
versus CERN-NA37 and FNAL-E665
data on DIS on nuclei. The band corresponds to the choice 
$\xi^2 = 0.09 - 0.12$~GeV$^2$. Data-Theory, where $\Delta_{D-T}$ 
is  computed for the set presented by circles, also shows comparison to the 
EKS98 scale-dependent shadowing parametrization.
Right panel from~\cite{Qiu:2003vd}: 
CERN-NA37 data on $F_2(Sn)/F_2(C)$ 
show evidence for a power-law  in $1/Q^2$ behavior consistent with 
the all-twist resummed calculation. The bottom right insert 
illustrates the role of higher twist 
contributions to $F_L$  on the example of $R=\sigma_L/\sigma_T$.  
}
\label{fig5-6}
\end{figure}

A class of corrections that can be naturally incorporated
in  the perturbative QCD factorization approach is associated 
with the power suppressed $\sim 1/Q^n,\; n \geq 1$ contributions.
The higher twist terms in Eq.~(\ref{series}) are typically 
neglected in reactions with ``elementary'' nucleons
for $Q^2 \geq 1$~GeV$^2$. However, in the presence of nuclear 
matter such corrections can be enhanced by the large nuclear 
size $\sim A^{1/3}$. These lead to dynamical nuclear shadowing. 
We discuss this  mechanism in detail below.

Hard scattering in nuclear collisions requires one large momentum
transfer $Q \sim xP \gg \Lambda_{QCD}$ with parton momentum 
fraction $x$ and beam momentum $P$. A simple example is the 
lepton-nucleus deeply inelastic scattering (DIS). The transverse 
area probed by the virtual meson $\sim 1/Q^2$ is very small for
moderate and large $Q^2$ processes. Hence, the struck quark 
will propagate in the nucleus and interact with nucleons along 
the same impact parameter.  If $Q^2 \ll  m_N^2$ the {\em transverse} 
size of the wave packet will be comparable or larger than the size 
of the nucleon. In  this case the rapid fall-off  of the parton wave 
function effectively limits $Q^2 \geq Q^2_0$.     
The effective longitudinal interaction length probed by the virtual meson
of momentum $q^\mu$ is characterized by $1/xP$. If the momentum fraction 
of an active initial-state parton $x \ll x_c=1/2m_N r_0\sim 0.1$  
with nucleon mass $m_N$ and radius  $r_0$, it could cover several  
Lorentz contracted nucleons of longitudinal size $\sim 2r_0 (m_N/P)$
in a large nucleus.

Each of the multiple soft interactions is characterized by a 
scale of power correction per nucleon   
\begin{equation}
\xi^2\approx \frac{3\pi\alpha_s(Q)}{8r_0^2}
\langle p|\hat{F}^2|p\rangle\;,
\label{scale}
\end{equation} 
with the matrix element  $\langle p|\hat{F}^2|p\rangle = 
\frac{1}{2} \lim_{x\rightarrow 0}xG(x,Q^2)$. A large nucleus will
provide $\sim A^{1/3}$ enhancement of the high twist shadowing 
contribution. We note the Eq.~(\ref{scale}) has the geometric 
average for minimum bias reactions incorporated in the 
definition of $\xi^2$.

\subsection{Applications of dynamical  power corrections}

In Ref.~\cite{Qiu:2003vd} we resummed the nuclear enhanced
high twist corrections and identified the modification to
the leading twist and lowest non-vanishing order 
in $\alpha_s$ contribution to the longitudinal and transverse 
structure functions: 
\begin{eqnarray} 
F_T^A(x,Q^2) & \approx & 
A \, F_T^{\rm (LT)}\left( x + \frac{x \xi^2 ( A^{1/3}-1) }{Q^2}, 
                      Q^2 \right) \, ,
\label{FTres}  \\ 
F_L^A(x,Q^2)   & \approx & 
A\,  F_L^{\rm (LT)}(x,Q^2) + 
\frac{ 4\, \xi^2 }{Q^2} \, F_T^A(x,Q^2)  \; .
\label{FLres}  
\end{eqnarray}
These results, when compared to Eq.~(\ref{strfun}), indicate 
that the essential difference from the final state coherent 
multiple parton scattering is the generation of dynamical mass
and the consequent rescaling in the value of Bjorken-$x$.
In addition, there is a novel contribution to 
$F_L^A$.

Calculations of nuclear shadowing versus $x_B$ 
with $\xi^2 = 0.09 - 0.12$~GeV$^2$ are 
given in left panel of  Fig.~\ref{fig5-6}. The right panel shows the $Q^2$ 
dependence of power corrections and the enhancement in the longitudinal
structure function, see Eq.~(\ref{FLres}), 
reflected in $R=\sigma_L/\sigma_T$.   
   
This resummation, derived within the framework of the pQCD 
factorization approach~\cite{Qiu:2003vd} has definite advantages:   

\begin{itemize}

\item  It relies on standard PDFs~\cite{Pumplin:2002vw} and its 
particularly easy to implement numerically. The final state 
soft scattering of the struck parton in the medium generates 
dynamical parton mass $m_{\rm dyn}^2 = \xi^2 ( A^{1/3} - 1 )$
via the coupling to the background chromo-magnetic field 
given by the two gluon correlation function, Eq.~(\ref{scale}). 
The necessary  additional energy leads to the rescaling 
of the value of Bjorken-$x$. This physical interpretation 
becomes transparent when we examine the behavior of 
$\nu+A$ DIS with charged 
current exchange and charm quark in the final state~\cite{Qiu:2004qk} 
with large physical mass $M_c$. Such processes are allowed via the 
CKM matrix mixing and we find: 
\begin{equation}
\hspace*{-1cm} x_B \rightarrow x_B + x_B \frac{M_c^2}{Q^2} +  
x_B \frac{\xi^2 (A^{1/3} -1)}{Q^2} = 
x_B \left( 1 + \frac{M^2_c + m_{\rm dyn}^2 }{Q^2}    \right)  \;.  
\label{StatDynShift} 
\end{equation}     
Similar rescaling of Bjorken-$x$ has been 
derived in~\cite{Mikkel} for the EMC effect.

\begin{figure}[t!]
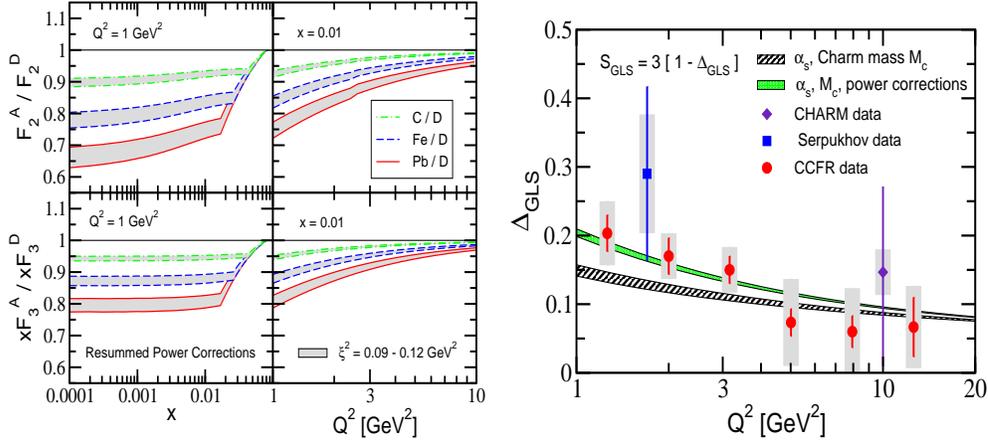

\includegraphics[width=2.5in,height=2.3in]{Neutrino-EMC.eps}
\hspace*{0.2cm}\includegraphics[width=2.5in,height=2.2in]{deltaGLS.eps} 
\caption{ Left panel from~\cite{Qiu:2004qk}: 
 predicted nuclear modification for isoscalar-corrected
$^{12}C, \, ^{56}Fe$ and  $^{208}Pb$ to the neutrino-nucleus 
DIS stricture functions $F^A_2(x_B,Q^2)$ (top) and $x_B F^A_3(x_B,Q^2)$ 
(bottom) versus Bjorken $x_B$ (left) and $Q^2$ (right). The bands 
correspond to $\xi^2 = 0.09-0.12$~GeV$^2$.
Right panel from~\cite{Qiu:2004qk}: 
 $\Delta_{\rm GLS}$ calculated to ${\cal O}(\alpha_s)$
with charm mass ($M_c=1.35$~GeV) effects (stripes) and $M_c$ +
resummed power corrections (band).
Data is from CCFR, CHARM and IHEP-JINR.
}
\label{fig7-8}
\end{figure}

\item  This resummation  provides a {\em natural framework} 
to understand the {\em differences} in the nuclear shadowing 
of sea quarks, valence quarks and gluons.  In the leading-order
and leading twist parton model in $\nu+A$ DIS 
$F_3^A(x_B,Q^2)$ measures the valance
quark number density with $\phi_{val.}(x) \propto x^{-\alpha_{val.}}$ 
at small~$x$. $F_2^A(x_B,Q^2)$, a singlet distribution, 
is proportional to the momentum density of all interacting 
quark constituents and for $x_B \ll 0.1$ is dominated by 
the sea contribution, $\phi_{sea}(x) \propto x^{-\alpha_{sea}}$.
Therefore, the $x_B$-dependent shift from dynamical nuclear 
enhanced power corrections, 
Eq.~(\ref{StatDynShift}), generates different
modification to $F_2^A(x_B,Q^2)$ and $F_3^A(x_B,Q^2)$. 
Predictions for the $x_B-$ and $Q^2-$dependence of shadowing 
in $\nu + A$ reactions is given in the left panel of Fig.~\ref{fig7-8}. 
The right panel  shows the  high-twist contribution.    
to the Gross-Llewellyn Smith QCD sum rule~\cite{Gross:1969jf}:
\begin{equation}
\hspace*{-1cm} S_{\rm GLS}  = 3(1-\Delta_{\rm GLS})  
= \int_0^1 dx_B \,\frac{1}{2x_B} 
\left( x_B F_3^{(\nu A)}+ x_B F_3^{(\bar{\nu} A)} \right)\; . 
%\nonumber 
\label{Sgls}
\end{equation}

\item  Gluon shadowing is also easily understood within the same
resummation approach as the final state scattering of the struck 
gluon~\cite{Qiu:2004da} in $p+A$ reactions. In the color 
singlet approximation gluons couple  twice as strongly 
to the medium in comparison to  quarks. The scale of power 
corrections then reads  
$(C_A/C_F) \xi^2 = 2.25 \xi^2 = 0.20 - 0.27$~GeV$^2$ and the 
corresponding dynamically generated gluon mass is twice as large.

\hspace*{.8cm} We now focus on the effects of dynamical nuclear shadowing 
on forward rapidity hadron production at RHIC. Any attenuation 
of the perturbative cross sections from coherent or inelastic 
parton scattering can be detected through the nuclear modification 
ratio $R^{(n)}_{AB}$, Eq.~(\ref{multi}). In the special case of 
dihadron correlations such cross section reduction will be
manifest in the attenuation of the area $A_{\rm far}$ of the
correlation function $C_2(\Delta \phi)$.
 In order to compute the process dependent nuclear shadowing 
we isolate the small $x_b$ dependence of  single and double
inclusive hadron production, Eqs.~(\ref{single}) and (\ref{double}), 
\begin{equation}
 F_{ab\rightarrow cd}(x_b) =   
\frac{\phi_{b/N}(x_b)}{x_b}\,  
|\overline {M}_{ab\rightarrow cd}|^2  \;. 
\end{equation}
The final state interactions of the struck parton lead to
rescaling of the small momentum fraction  $x_b$ as 
follows~\cite{Qiu:2004da}: 
\begin{equation}
 F_{ab\rightarrow cd}(x_b)  \Rightarrow  
 F_{ab\rightarrow cd} \left(x_b \left[ 1+ C_d
\frac{\xi^2}{-t} (A^{1/3}-1) \right]\right)  \;.
\label{resum-t}
\end{equation}
In Eq.~(\ref{resum-t}) $C_d$ keeps track of the representation
of the struck parton with $C_d = 1$ for quarks and 
$C_d =  2.25$ for gluon, as discussed above.

\begin{figure}[t!]
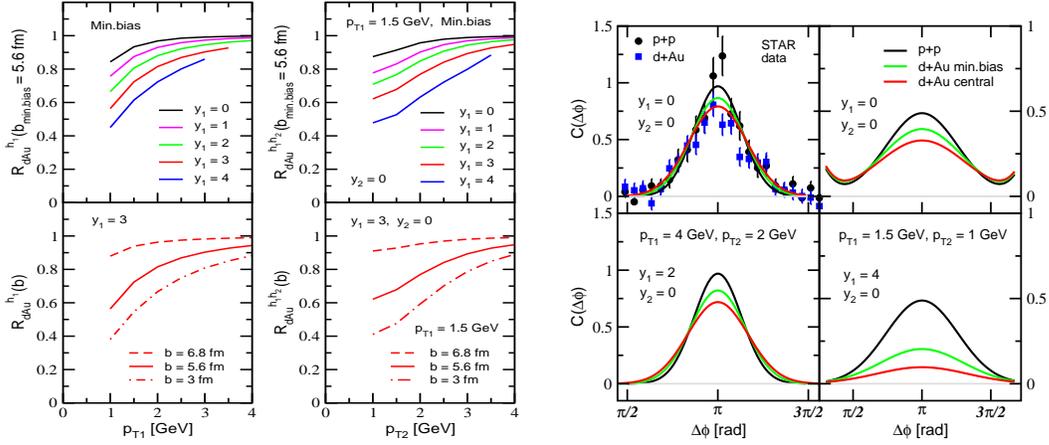

\includegraphics[width=2.7in,height=2.3in]{pA-G-Shadow.eps}
\hspace*{0.5cm}
\includegraphics[width=2.5in,height=2.2in]{pA-Correlation.eps} 
\caption{ Left panel from~\cite{Qiu:2004da}: 
suppression of the single and  double inclusive 
hadron production rates in d+Au reactions versus $p_T$ for 
rapidities $y_1 = 0,1,2,3$ and $4$. $\xi^2 = 0.12$~GeV$^2$.  
Also shown is the impact parameter dependence of the calculated 
nuclear modification for central $b=3$~fm,
minimum bias $b_{\rm min.bias}=5.6$~fm and peripheral
$b=6.9$~fm collisions. The trigger hadron $p_{T1}=1.5$~GeV,   
$y_1 = 3$ and the associated hadron  $y_2=0$.
Right panel from~\cite{Qiu:2004da}: centrality dependence of 
$C_2(\Delta \varphi)$ at moderate  $p_{T_1}=4$~GeV, 
$p_{T_2}=2$~GeV and rapidities $y_1=0,2$ and $y_2=0$.  
Central d+Au and p+p data  from STAR.
Also shown is $C_2({\Delta \varphi})$ at small transverse momenta 
$p_{T_1}=1.5$~GeV, $p_{T_2}=1$~GeV and rapidities $y_1=0,4$ 
and $y_2=0$.}
\label{fig9-10}
\end{figure}

Figure~\ref{fig9-10} show the predicted suppression of the 
single and  double inclusive differential hadron production 
cross sections versus rapidity and centrality. The scale 
of power corrections
was extracted for minimum bias collisions and  the impact 
parameter dependence can be obtained via the corresponding 
rescaling with the nuclear thickness function. It is important
to note that the value of $\xi^2$ remains {\em unchanged}. 
However, in going from midrapidity to forward rapidity
the perturbative scale $-t$ that enters in Eq.~(\ref{resum-t})
changes from $\sim 2p_T^2$ to   $\sim  p_T^2$. This trivial 
kinematic dependence, which leads to larger suppression
at forward rapidity,  has been neglected in previous 
calculations. The right panel of Fig.~\ref{fig9-10} 
shows  the combined effect of the the small transverse 
momentum  broadening and nuclear enhanced power corrections. 
In contrast to alternative hypotheses, the high-twist 
shadowing goes away quickly with the virtuality 
or transverse momentum as long as the large $x_a \rightarrow 1$ 
threshold effects remain unimportant~\cite{Kopel}.

Further observations on the effects of  nuclear shadowing 
on hadron production in $d+Au$ at RHIC is available 
in~\cite{Vogt:2004cs}. Preliminary study of the 
effects of higher-twist 
resummation~\cite{Qiu:2003vd,Qiu:2004qk,Qiu:2004da}
on the low $Q^2$ modification to the DGLAP evolution equations
has recently become available~\cite{Zhu:2004zw}.

\end{itemize}

\section{Jet quenching and jet tomography}

\begin{figure}[t!]
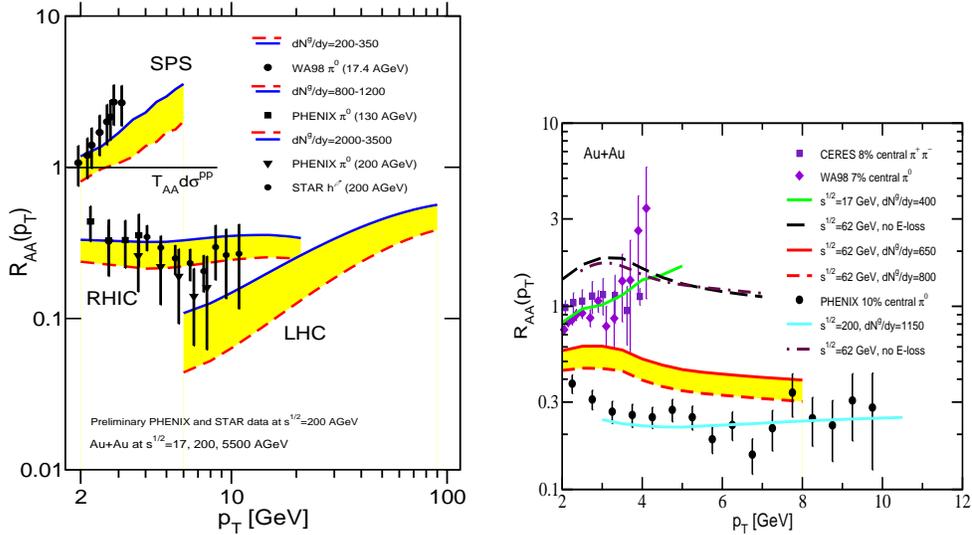

\includegraphics[width=2.4in,height=2.8in]{Quench-200GeV.eps}
\hspace*{0.5cm}\includegraphics[width=2.4in,height=2.2in]{Quench-62GeV.eps} 
\caption{ Left panel from~\cite{Vitev:2002pf}:  
suppression/enhancement ratio $R_{AA}(p_T)$  
    for neutral 
pions at $\sqrt{s}_{NN}=17$, $200$, $5500$~GeV.   Solid (dashed) 
         lines correspond to the smaller (larger)  effective initial  
         gluon rapidity densities at given $\sqrt{s}$ that drive parton 
         energy loss. 
Right panel from~\cite{Vitev:2004gn}: predicted nuclear modification 
factor at $\sqrt{s_{NN}} = 62 $~GeV for central $Au+Au$ collisions.
Enhancement, arising from 
transverse momentum diffusion in cold nuclear matter 
without final state energy loss
is given for comparison.
}
\label{fig11-12}
\end{figure}

\begin{figure}[t!]
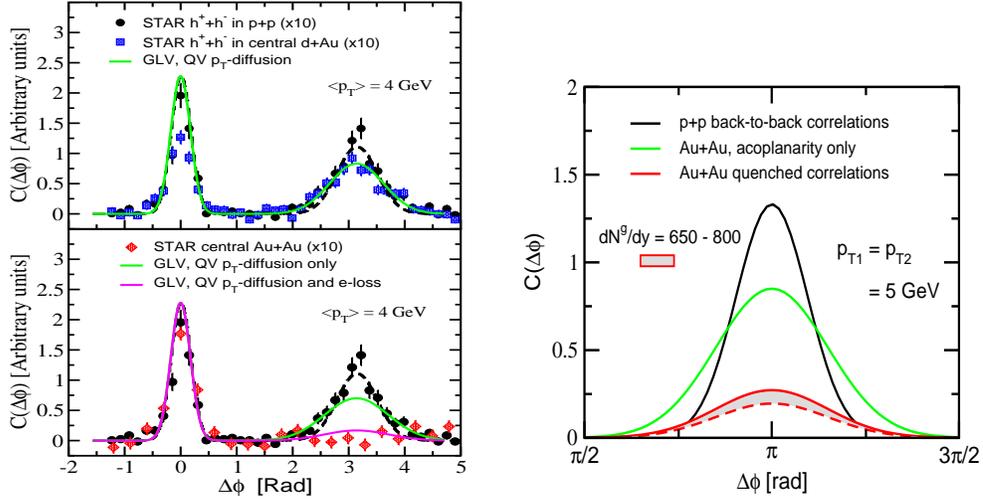

\includegraphics[width=2.4in,height=2.6in]{Away-Correlation-200GeV.eps}
\hspace*{0.5cm}
\includegraphics[width=2.4in,height=2.2in]{Away-Corr-62GeV.eps} 
\caption{ Left panel from~\cite{Vitev:2003jg}:
the broadening of the 
away-side dihadron correlation function in central $d+Au$ and $Au+Au$ 
compared to scaled  STAR data. In the bottom  panel 
the broadening with and without suppression, $\propto R_{AA}$, 
is shown. Right panel from~\cite{Vitev:2004gn}: attenuation of the 
double inclusive pion production cross section for 
$p_{T1} = p_{T2} = 5$~GeV at intermediate RHIC energies.
}
\label{fig13-14}
\end{figure}

One of the few predicted~\cite{Wang:1991xy} and observed 
signatures of nuclear dynamics associated with the presence 
of a hot and dense QCD plasma is the suppression of the 
high-$p_T$ hadrons  or jet quenching. While different 
attenuation  mechanisms have been proposed~\cite{Gallmeister:2004iz},
it is the multiple inelastic parton scattering in
the final state that was able to quantitatively predict
many of the hadronic observables associated with  jet 
quenching~\cite{Adler:2002tq,d'Enterria:2004nv,Mioduszewski:2004nc}. 
Different approximations to the medium-induced 
non-Abelian gluon bremsstrahlung dynamics can be 
found in~\cite{Gyulassy:2003mc}. In this section we give 
details of the Reaction Operator approach~\cite{Gyulassy:2000er},  
a momentum space iterative scheme for computing the radiation 
intensity.

The full solution for the medium induced gluon radiation
off jets produced in a hard collisions at early 
times $\tau_{jet} \simeq 1/E$ inside a  nuclear  
medium  of length $L$  can be obtained to all orders  
in the correlations between the multiple scattering  
centers in the GLV approach~\cite{Gyulassy:2000er}.
The double differential bremsstrahlung  intensity 
for gluons with momentum $k=[xp^+, {\bf k}^2 / xp^+,{\bf k}]$  
resulting from the sequential interactions 
of a fast parton with momentum $p=[p^+, 0,{\bf 0}]$ 
can be written as: 
\begin{eqnarray}
\hspace*{-2.5cm} x\frac{dN_g}{dx\, d^2 {\bf k}}  &=&
\sum\limits_{n=1}^\infty  x\frac{dN_g^{(n)}}{dx\, d^2 {\bf k}}   
 = \sum\limits_{n=1}^{\infty}  \frac{C_R \alpha_s}{\pi^2} 
 \; \prod_{i=1}^n \;\int_0^{L-\sum_{a=1}^{i-1} \Delta z_a } 
 \frac{d \Delta z_i }{\lambda_g(i)}  \, \nonumber \\[1.ex]   
\hspace*{-1cm} &\times& \int   d^2{\bf q}_{i} \, 
\left[  \sigma_{el}^{-1}(i)\frac{d \sigma_{el}(i)}{d^2 {\bf q}_i}
  - \delta^2({\bf q}_{i}) \right]  \,  
\left( -2\,{\bf C}_{(1, \cdots ,n)} \cdot 
\sum_{m=1}^n {\bf B}_{(m+1, \cdots ,n)(m, \cdots, n)}   \right. 
\nonumber \\[1.ex] 
\hspace*{-1cm}&\times& \left.  \left[ \cos \left (
\, \sum_{k=2}^m \omega_{(k,\cdots,n)} \Delta z_k \right)
-   \cos \left (\, \sum_{k=1}^m \omega_{(k,\cdots,n)} \Delta z_k \right)
\right]\; \right) \;, \quad \qquad  
\label{difdistro} 
\end{eqnarray}
where $\sum_2^1 \equiv 0$ is understood. In the small angle 
eikonal limit $x=k^+/p^+ \approx \omega/E$. In Eq.~(\ref{difdistro})  
the color current propagators, the formation times and the elastic
scattering cross section are defined in~\cite{Gyulassy:2000er}.
For a derivation in the case of heavy quarks 
see~\cite{Djordjevic:2003zk}.

Qualitatively, the behavior of the energy loss  as a function 
of the density and the size of the system can be summarized 
to first order in opacity as follows:
\begin{eqnarray}
\langle \Delta E \rangle &\approx&  \int dz \, 
\frac{C_R\alpha_s}{2} \frac{\mu^2}{\lambda_{g}} \, z \, 
\ln \frac{2E}{\mu^2 \langle L \rangle }   
= \int dz \,  \frac{9 C_R \pi \alpha_s^3}{4} \rho^g(z) 
\, \ln \frac{2E}{\mu^2 \langle L \rangle }  \nonumber  \\[1ex]
&& \qquad \quad  \!\! = \left\{ \begin{array}{ll}  
\frac{9 C_R \pi \alpha_s^3}{8} \, \rho^g \langle L \rangle^2 
\, \ln \frac{2E}{\mu^2 \langle L \rangle } 
 \, ,  &  {\rm static} \\[1ex]
 \frac{9  C_R \pi \alpha_s^3}{4} 
\frac{1}{A_\perp}  \frac{dN^{g}}{dy} \langle L  \rangle 
\,   \ln \frac{2E}{\mu^2 \langle L \rangle}  
 \, , & 1+1D  
\end{array}  \right.   .
\label{deltae}
\end{eqnarray}
For static systems $\langle \Delta E \rangle$ depends 
quadratically on the nuclear size. For the case of longitudinal 
Bjorken expansion this dependence is reduced to 
linear~\cite{Gyulassy:2000gk}
but  the energy loss is sensitive to the initial parton rapidity 
density $dN^g/dy$. Understanding the effective color charge 
density dependence  of  $\Delta E$  the  is the key to 
jet tomography~\cite{Vitev:2002pf}. Transverse expansion does not affect 
significantly the attenuation of the inclusive 
spectra~\cite{Gyulassy:2000gk}.

\begin{figure}[t!]
\includegraphics[width=2.7in,height=2.3in]{Vitev-Reappear.eps}
\hspace*{0.5cm}\includegraphics[width=2.3in,height=2.in]{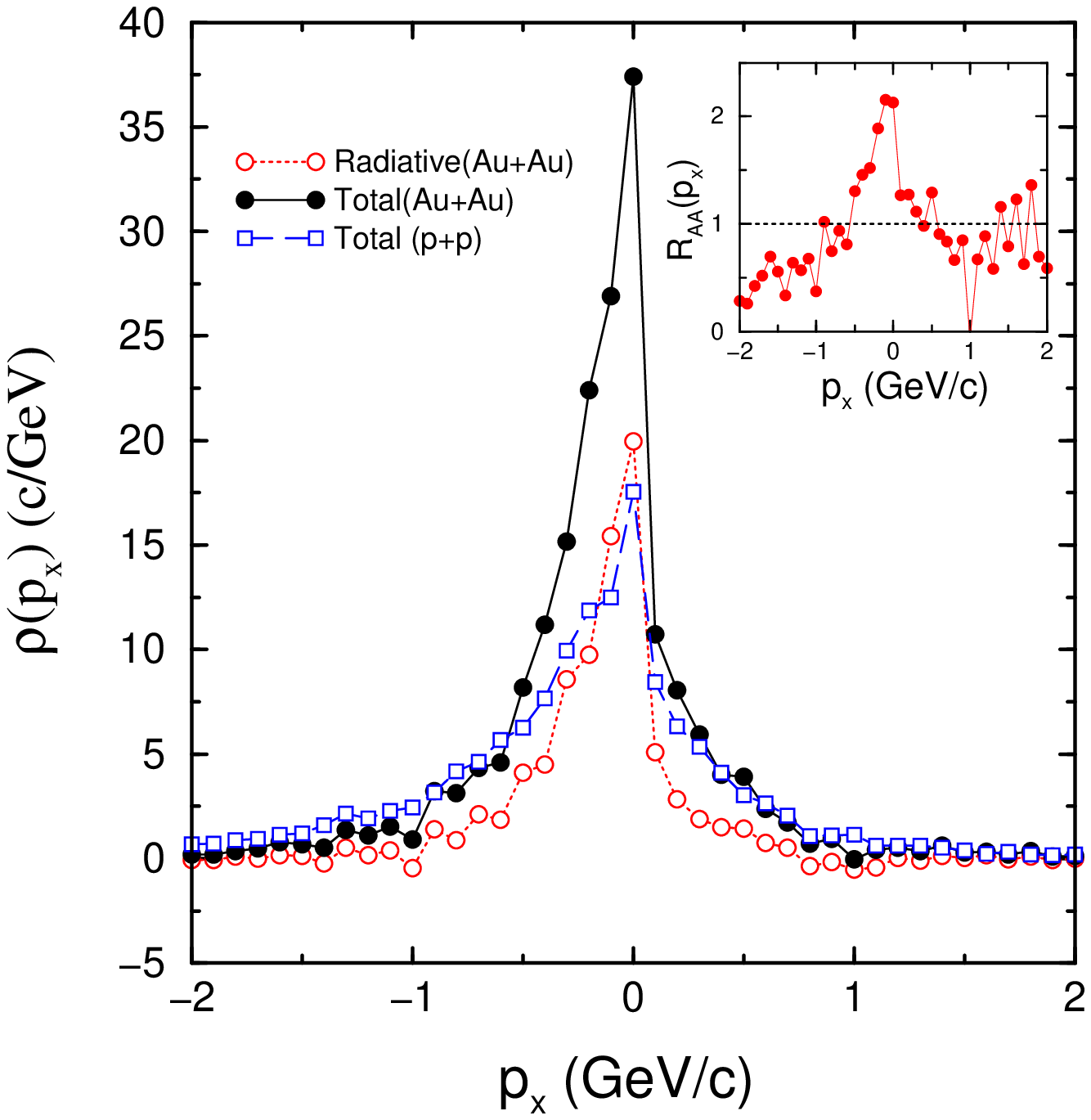} 
\caption{ Left panel from~\cite{Vitev:2002pf}: 
induced partonic multiplicity as a function of the 
experimental $p_{T\, {\rm cut}}$ for energetic quark jets at RHIC and 
the LHC. Right panel from~\cite{Pal:2003zf}: momentum density 
of hadrons associated with  energetic back-to-back jets with and 
without medium induced bremsstrahlung. Secondary rescattering leads
to gluon transverse momenta $\sim 600$~MeV. }
\label{reapp}
\end{figure}

\subsection{Applications of jet quenching}

The probability $P(\epsilon)$ for fractional energy loss  
$\epsilon = \sum_i \omega_i / E$ due to multiple gluon emission is 
evaluated in the independent Poisson approximation~\cite{Baier:2001yt}. 
One way of implementing the parton momentum attenuation  is via 
the kinematic rescaling of the momentum fraction 
$z_j \rightarrow z_j / (1-\epsilon)$ in the decay 
functions, Eq.~(\ref{factorization}). Consequently,
\begin{equation}
D_{h_j/j}(z_j, \mu_{d\, j}) \rightarrow 
 \int d\epsilon  \;  P(\epsilon)  \frac{1}{1-\epsilon} D_{h_j/j} 
\left( \frac{z_j}{1-\epsilon}, \mu_{d\, j} \right) \;. 
\label{D-rescale}
\end{equation}   
These corrections to the factorized formulas, Eqs.~(\ref{single}) 
and (\ref{double}), can be easily implemented numerically.

The left panel of Fig.~\ref{fig11-12} shows the 
predicted nuclear modification~\cite{Vitev:2002pf}
as a function of $\sqrt{s_{NN}}$ at SPS, RHIC and the LHC as a
function of the density of the quark-gluon plasma. The $p_T$ 
dependence of $R^{(1)}_{AA}$ is a result of the interplay of 
the Cronin effect, jet quenching and nuclear 
shadowing~\cite{Eskola:1998df}. 
The right panel shows a theoretical estimate
for the  $\pi^0$ (or $\pi^+ + \pi^-$)  attenuation at the 
intermediate RHIC energy of  
$\sqrt{s_{NN}}=62$~GeV~\cite{Vitev:2004gn}. 
Such suppression is in agreement with the preliminary PHENIX and 
STAR measurements~\cite{d'Enterria:2004nv,prel-62} 
for $ p_T > 3$~GeV. It is interesting to observe the better 
agreement between data and theory at the SPS with a 
recently extracted low energy $p+p$ 
baseline~\cite{d'Enterria:2004ig}. 
Similar suppression at the intermediate RHIC
energy  has been found in~\cite{Adil}.

For double inclusive hadron production  qualitatively 
$1 \leq R^{h_1}_{AA} / R^{h_1h_2}_{AA} \leq 2$. This attenuation
will be manifest as a reduction of the area  $A_{\rm Far}$ of 
$C_2(\Delta \phi)$, Eq.~(\ref{cor-fun}). The left panel of 
Fig.~\ref{fig13-14} demonstrates that transverse momentum 
diffuion in $Au+Au$ collisions is insufficient to reproduce
the dissapearance of the away-side correlations~\cite{Adler:2002tq}. 
Comparison to data require significant jet energy loss, compatible 
with the attenuation in the single inclusives. 
The right panel shows the predicted quenching effect at 
$\sqrt{s_{NN}}=62$~GeV~\cite{Vitev:2004gn}.

The energy lost by the energetic partons during their propagation
in dense nuclear matter is redistributed back in the partonic system. 
In the GLV approach~\cite{Gyulassy:2000er} the medium induced 
virtuality is irradiated in higher frequency modes  
$\omega  \geq  \omega_{\rm pl} \sim \mu $ resulting in fewer more 
energetic gluons. The left panel of Fig.~\ref{reapp} shows the 
the recovered jet energy as a function of the experimental 
$p_{T \; \rm cut}$  and the corresponding  
induced parton multiplicity + 1(jet). 
If the secondary (bremsstrahlung) gluons reinteract in the 
system~\cite{Pal:2003zf} 
their momentum is further degraded to $p_T \simeq 600$~MeV and 
\begin{equation}
N_g({\bf r}, \Delta\tau) \approx \frac{1}{4}\Delta S =
\frac{1}{4} \frac{\Delta E({\bf r},\Delta\tau)} {T({\bf r},\tau)} ~.
\label{ngluon}
\end{equation}
Numerical results for this scenario are shown in the 
right hand side of Fig.~\ref{reapp}.
For preliminary results on the recovery of the lost energy 
in jet measurements see~\cite{Adler:2002tq,Wang:2004kf}.

\section{Conclusions} 

The predictive power of perturbative QCD is based one
the factorization approach that has been successfully tested 
in a variety of moderate and high transverse momentum 
processes~\cite{Collins:gx,Collins:1981uw,Collins:1981uk,Bodwin:1984hc,Ellis:1978sf,Bodwin:1994jh} 
in elementary $N+N$ collisions.  
In this talk I discussed the theoretical and phenomenological 
aspects of the systematic incorporation of nuclear enhanced 
higher order and higher twist corrections in this formalism 
in $\ell + A$,   $p+A$ and $A+A$  reactions.
The current data on relativistic heavy ion reactions 
and the extended  kinematic reach that is expected to 
become available due to 
improved statistics, RHIC upgrades, RHIC II and the LHC 
will help to further quantify the relative importance 
of the elastic~\cite{Qiu:2003pm,Accardi:2002ik}, 
inelastic~\cite{Gyulassy:2003mc,Gyulassy:2000er} 
and coherent~\cite{Qiu:2003vd,Qiu:2004qk,Qiu:2004da} 
multiple parton scattering  in cold and hot nuclear matter.

\vspace*{.3cm}

\noindent {\bf Acknowledgments: } 
I would like to thank Mikkel Johnson for useful discussion. This
work is supported by the J.R.~Oppenheimer Fellowship 
of the Los Alamos National Laboratory and by the US Department of Energy.

\end{document}